\documentclass[aps, prb, twocolumn, showpacs, floatfix, amsmath, amssymb, reprint]{revtex4-1}

\usepackage{graphicx}
\usepackage{dcolumn}			
\usepackage{bm}				
\usepackage{hyperref}
\usepackage{psfrag}
\begin{document}

\title{Bulk sensitive x-ray absorption spectroscopy free of self-absorption effects}

\author{A. J. Achkar$^1$, T. Z. Regier$^2$, H. Wadati$^3$, Y.-J. Kim$^4$, H. Zhang$^4$ and D. G. Hawthorn$^1$ }
\affiliation{$^1$Department of Physics and Astronomy, University of Waterloo, Waterloo, N2L 3G1, Canada \\ $^2$Canadian Light Source, University of Saskatchewan, Saskatoon, Saskatchewan S7N 0X4, Canada \\ $^3$Department of Physics and Astronomy, University of British Columbia, Vancouver, British Columbia V6T 1Z1, Canada \\ $^4$Department of Physics, University of Toronto, 60 St. George Street, Toronto, Ontario M5S 1A7, Canada}

\date{\today}

\begin{abstract}
We demonstrate a method of x-ray absorption spectroscopy (XAS) that is bulk sensitive, like traditional fluorescence yield measurements, but is not affected by self-absorption or saturation effects. This measure of XAS is achieved by scanning the incident photon energy through an absorption edge and using an energy-sensitive photon detector to measure the partial fluorescence yield (PFY). The x-ray emission from any element or core-hole excitation that is not resonant with the absorption edge under investigation is selected from the PFY.  It is found that the inverse of this PFY spectrum, which we term inverse partial fluorescence yield (IPFY), is linearly proportional to the x-ray absorption cross-section without any corrections due to saturation or self-absorption effects. We demonstrate this technique on the Cu $L_{2,3}$ and Nd $M_{4,5}$ absorption edges of the high-$T_c$ cuprate $\rm {La_{1.475}Nd_{0.4}Sr_{0.125}CuO_4}$ by measuring the O $K_{\alpha}$ PFY and comparing the total electron yield, total fluorescence yield, and IPFY spectra. 
\end{abstract}

\maketitle

\pdfbookmark[1]{Introduction}{Introduction}
Since the development of brilliant sources of synchrotron radiation, x-ray absorption spectroscopy (XAS) has become a widely used probe of the electronic and spatial structure of materials in various fields of the natural sciences. Multiple beamlines dedicated to XAS have been developed at virtually every synchrotron facility in the world. XAS is a core-level spectroscopy where a photon excites a tightly bound core electron above the Fermi energy into an unoccupied state. In the near-edge region (XANES), the resulting spectra provide element- and orbital-specific information about the unoccupied states of a material, revealing valuable information, such as the valence, spin-state, orbital symmetry, crystal field interaction, and the hybridization of atoms with their neighbors.\cite{deGroot08,Stohr96,deGroot94} Further above an x-ray absorption edge, oscillations in the absorption coefficient (EXAFS) provide information about the local atomic structure.\cite{Wende04}

X-ray absorption spectroscopy is most often measured in one of three ways: in transmission, using the electron yield (EY),\cite{Gudat72} or using the fluorescence yield (FY).\cite{Jaklevic77}  Transmission is perhaps the most direct method of measuring absorption.  However, it is not commonly used in the soft x-ray region since the rapid attenuation of x-rays in materials limits transmission measurements to ultra-thin samples ($\sim $1000 \AA), which is impractical for most studies.  More common are EY and FY measurements.  These techniques operate on the principle that the absorption cross-section is proportional to the number of core-holes created.  These core holes are filled by an electron having a lower binding energy, thereby emitting photons or electrons as decay products.  By measuring the decay products one can achieve a measure of the linear absorption coefficient.  Of these two measures, EY is more surface sensitive, with the depth sensitivity given by the escape depth of electrons [20--200~\AA (Refs. 1-3)], whereas FY is more bulk sensitive, with the depth sensitivity given by the penetration depth of the incident and emitted photons (of order 1000 \AA ~in the soft x-ray region).  FY is commonly measured as either the total fluorescence yield (TFY), which sums over all or a broad range of photon energies, or as the partial fluorescence yield (PFY), which uses an energy-sensitive detector to select fluorescence from a range of emitted photon energies.  PFY has been used as a probe which is sensitive to dilute concentrations of an element by focusing on the resonant emission from that element.\cite{Jaklevic77}  

Restricting the use of FY, however, are self-absorption effects \cite{Eisebitt93,Troger92} which suppress the peaks in the measured spectra and make FY nonlinear with respect to the absorption coefficient. In FY, self-absorption effects occur when the penetration depth of the x-rays varies strongly near an x-ray absorption edge. EY can also exhibit distortions, although they are typically less significant than the self-absorption effects in FY. In EY, a saturation effect occurs when the electron escape depth is comparable to the x-ray penetration depth \cite{Nakajima99} (at its extreme all incident photons will generate electrons that escape the sample, producing an energy-independent signal). In transmission, distortions can occur due to pinholes in samples or another saturation effect which comes from the exponential dependence of the transmission intensity on the absorption coefficient, limiting the dynamic range of transmission and leading to thickness-dependent distortions of the measured spectra.\cite{Esteva83}

In this paper, we introduce a method to measure the absorption coefficient that is bulk sensitive and does not suffer from saturation or self-absorption effects. The approach uses an energy-sensitive detector to measure the PFY. Unlike previous PFY measurements,\cite{Jaklevic77,Eisebitt93} however, we measure the normal (i.e., nonresonant) x-ray emission spectrum (NXES) from a different element than the absorption edge we are probing with incident photons. The inverse of this NXES PFY, which we term inverse partial fluorescence yield (IPFY), is shown to be experimentally (at the Cu $L$ and Nd $M$ edges of $\rm{La_{1.475}Nd_{0.4}Sr_{0.125}CuO_4}$) and theoretically proportional to the absorption cross-section and to be free of self-absorption or saturation effects.  The principle of the technique is different than standard FY or EY.  While FY and EY are measures of the number of core-hole excitations through an absorption edge, IPFY is effectively a measure of the x-ray attenuation length, akin to transmission measurements.  This  technique is anticipated to have wide-reaching applicability to XAS in a variety of materials using either hard or soft x-rays.

\begin{figure}[hb]
\begin{center}
\includegraphics[width = \columnwidth]{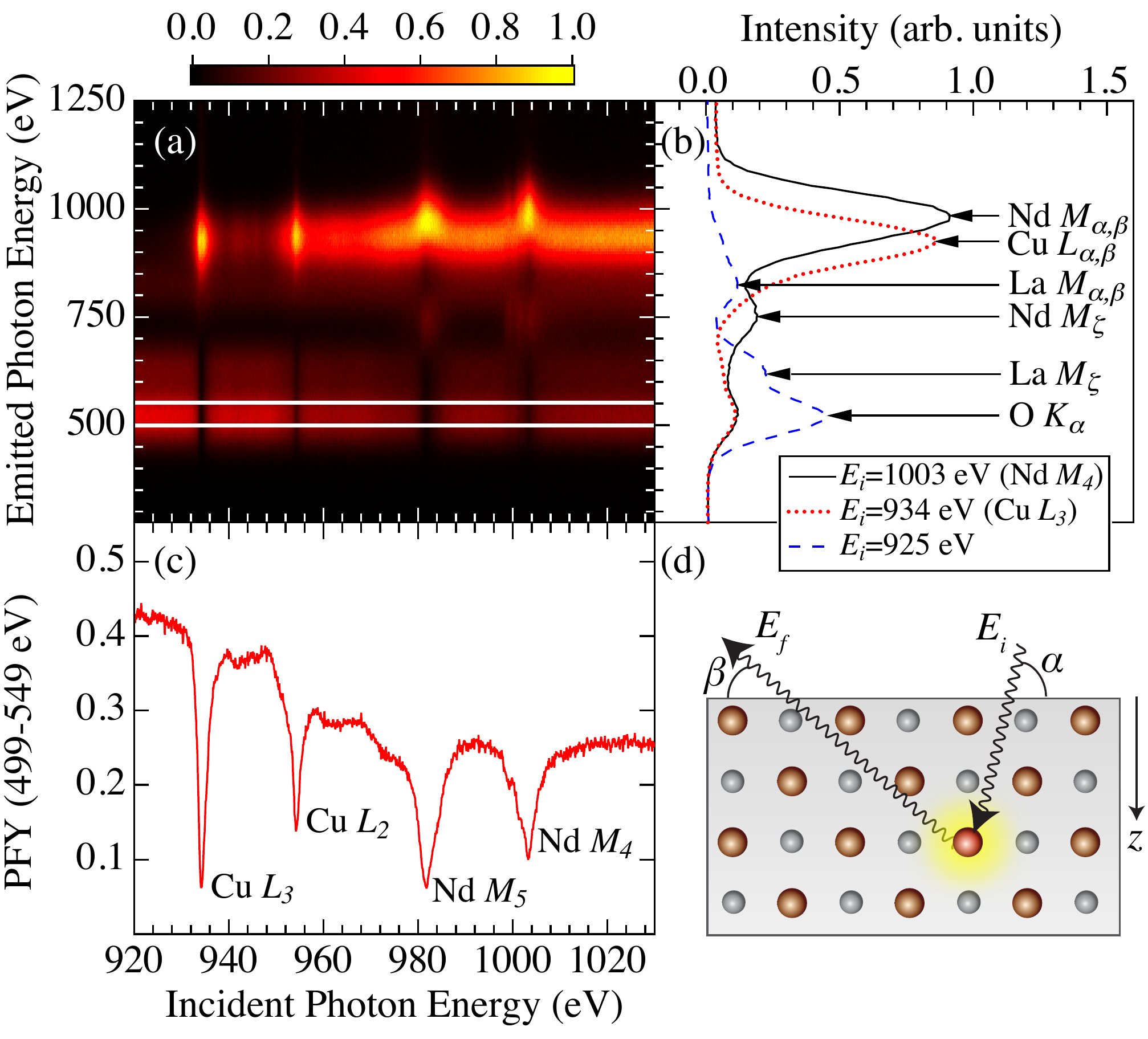}
\vspace{-15pt}
\caption{(Color online) (a) Normalized PFY of LNSCO as a function of $E_i$ and $E_f$ as the incident photon energy is scanned through the Cu $L$ and Nd $M$ edges.  (b) The emission spectra for incident photon energies of 1003 eV (black), 934 eV (dotted red), and 925 eV in (dashed blue) taken in a 1-eV window. Emissions corresponding to the Nd $M_{\alpha,\beta}$, Cu $L_{\alpha, \beta}$, La  $M_{\alpha, \beta}$, Nd $M_\zeta$, La $M_\zeta$, and O $K_{\alpha}$ transitions are observed as a function of emitted photon energy.  (c) The PFY at the O $K$ edge integrated over a 50-eV window centered at 524 eV [the region spanned by the horizontal lines in (a)].  The O $K$ edge emission dips at the Cu $L$ and Nd $M$ absorption edges where the absorption is maximal. (d) The experimental geometry: x-rays of energy $E_i$ are incident on the sample at an angle of $\alpha$ relative to the sample surface. Incident x-rays excite a core electron in an atom (highlighted in gold).  This atom decays back to the ground state by emitting a photon of energy $E_f$ that is detected at angle $\beta$ relative to the surface.}
\label{fig:fig1}
\end{center}
\end{figure}

\begin{figure}[hb]
\begin{center}
\includegraphics[width=\columnwidth]{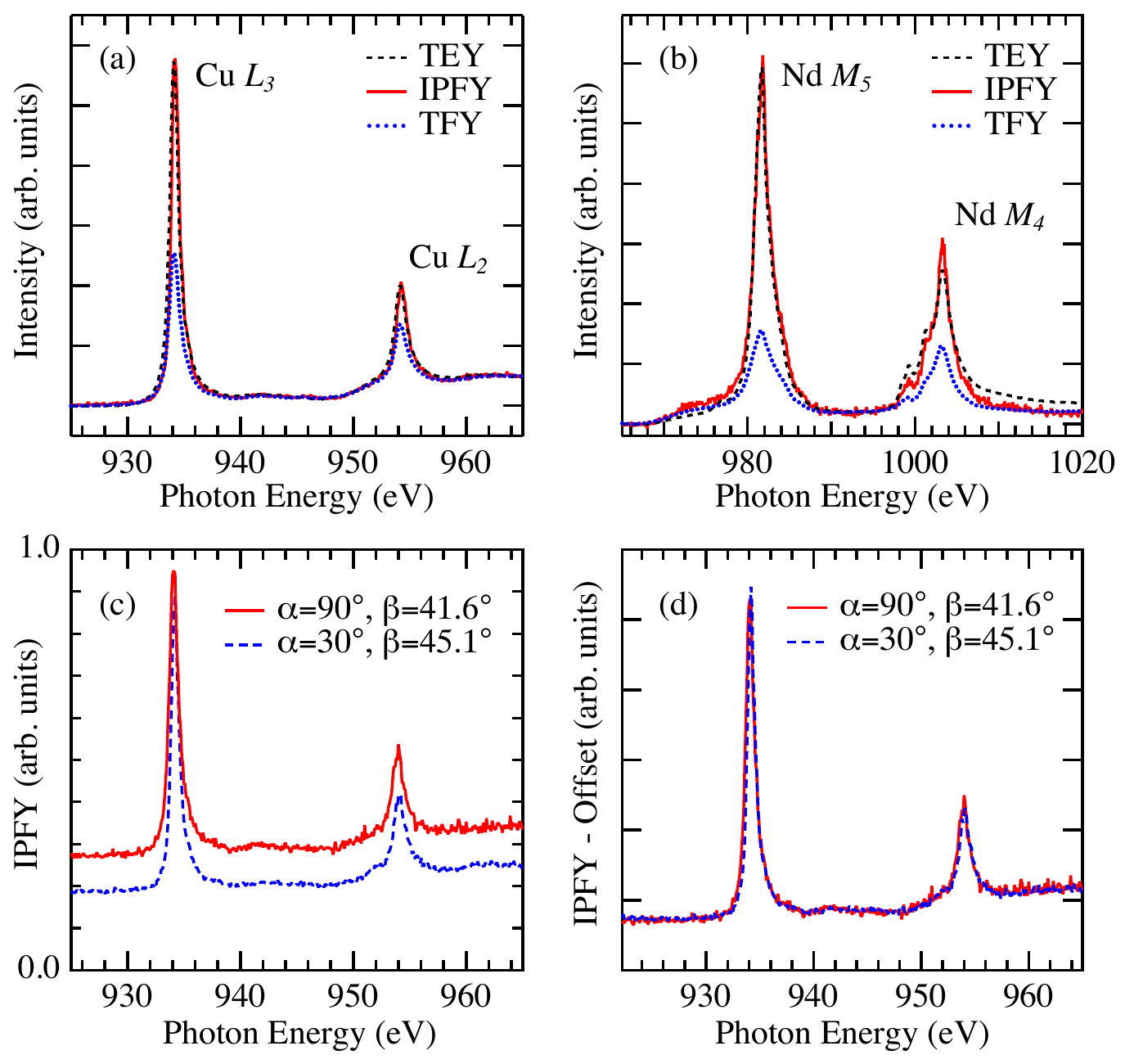}
\vspace{-20pt}
\caption{(Color online) The TEY (dashed black), O $K$ edge IPFY (red) and TFY (dotted blue) as a function of photon energy through the (a) Cu $L$ and (b) Nd $M$ edges.  The spectra are scaled and offset to match above and below the absorption edges.  The IPFY and TEY spectra are in good agreement while the TFY spectra are distorted due to self-absorption effects and NXES contributions to the TFY. (c) The Cu $L$ IPFY measured for two different geometries. (d) The IPFY spectra offset to match at a point in the pre-edge region.}
\label{fig:fig2}
\end{center}
\end{figure}

\pdfbookmark[1]{Experimental}{Experimental}
A single crystal of $\rm{La_{1.475}Nd_{0.4}Sr_{0.125}CuO_4}$ (LNSCO), grown by the traveling solvent floating zone method, was studied at the Canadian Light Source's 11ID-1 SGM beamline. The LNSCO crystal was cleaved along the $a-b$ plane in vacuum at a pressure of $\rm5\cdot10^{-8}$ Torr and maintained at this pressure during measurement. The total electron yield (TEY) was measured using drain current and the total fluorescence yield was measured with a channel plate detector.  The energy resolved PFY was also measured using a silicon drift detector with an energy-resolution of $\sim$100 eV. The sample temperature was $\sim$200K for data in Figs. \ref{fig:fig1}, \ref{fig:fig2}(a) and \ref{fig:fig2}(b) and room temperature for the data in Figs. \ref{fig:fig2}(c) and \ref{fig:fig2}(d).  The sample geometry is indicated in Fig. ~\ref{fig:fig1}(d).  For the measurements in Figs. ~\ref{fig:fig1} and~\ref{fig:fig2}(a) and \ref{fig:fig2}(b), photons were set normal to the sample surface ($\alpha\!=\!90^{\circ} $) and the channel plate and silicon drift detectors were positioned at $\beta\!\simeq\!52^{\circ}$ and $41.6^{\circ}$, respectively. For all measurements, the polarization was linear horizontal and perpendicular to the sample's $c$ axis. 

In Fig.~\ref{fig:fig1} we show the energy-resolved x-ray emission (normalized to the incident photon intensity) as the incident photon energy is scanned through the Cu $L$ and Nd $M$ absorption edges.  As a function of emitted photon energy, x-ray emission peaks are observed corresponding to the following transitions Nd $M_{\alpha,\beta} \simeq 990$ eV $(4f \rightarrow 3d)$, Cu $L_{\alpha, \beta} \simeq 940$ eV $(3d \rightarrow 2p)$, La  $M_{\alpha, \beta} \simeq 830$ eV $(4f \rightarrow 3d)$, Nd $M_\zeta \simeq 760$ eV $(4p \rightarrow 3d)$, La $M_\zeta \simeq 630$ eV $(4p \rightarrow 3d)$ and O $K_\alpha \simeq 524$ eV $(2p \rightarrow 1s)$.  As expected, the Cu $L$ and Nd $M$ emission peaks as the incident energy passes through the Cu $L$ and Nd $M$ edges, respectively.  In contrast, the O $K$ emission dips at the Cu $L$ and Nd $M$  absorption edges, as shown in Fig.~\ref{fig:fig1}(c) by the integrated intensity of the O $K$ emission vs. incident photon energy.  

The latter result can easily be understood: As the absorption increases (penetration depth decreases) through the Cu $L$ and Nd $M$ edges, fewer oxygen atoms are excited and subsequently fewer oxygen atoms emit at the O $K$ edge.  Moreover, it is apparent that the inverse of the O~$K$ emission, which we label IPFY, is proportional to the x-ray absorption and can be a useful measure of XAS.  This is verified by comparing, in Fig.~\ref{fig:fig2}, the IPFY signal to the TEY and TFY, which agree with previous work on XAS of LSCO \cite{Chen92} and Nd.\cite{Thole85}  The TEY and IPFY agree well for the Cu $L$ and Nd $M$ edges, although some discrepancy is seen at the Nd $M_{4}$ edge.  In contrast, the TFY is influenced by self-absorption, which visibly distorts the spectra. The IPFY measurement, however, is not distorted by self-absorption effects and provides an absorption spectrum comparable in shape to TEY while maintaining a probing depth similar to TFY.

\pdfbookmark[1]{Theory}{Theory}
These experimental findings can be understood by examining the expected x-ray emission intensity $I(E_i, E_f)$ as a function of the incident and emitted photon energies, $E_i$ and $E_f$ respectively.  $I(E_i, E_f)$ normalized to the incident photon intensity, $I_0(E_i)$, is given by
\begin{equation} 
	 	\frac{I(E_i, E_f)}{I_0(E_i)} \!=\! C \! \sum_{X} \!\! \displaystyle\int^\infty_0\!\!\!\!\! \omega_{X}\!(E_i, E_f )\frac{\mu_{X}\!(E_i)}{\sin \alpha}e^{- \! \left (\! \frac{\mu(E_i)}{\sin \alpha} + \frac{\mu(E_f)}{\sin \beta}\! \right )z}   \!\! \ dz, 
\label{eqn:step1}
\end{equation}
where $C\!=\!\eta(E_f)\Omega/4\pi$ is a constant given by the detector efficiency $\eta(E_f)$ and the solid angle $\Omega$ of the detector.  $\mu(E_i)$ and $\mu(E_f)$ are the total linear attenuation coefficients of the material at energies $E_i$ and $E_f$. $\alpha$ and $\beta$ are the angle of incidence and angle of detection, respectively.   The exponential terms in eqn.~\ref{eqn:step1} account for the attenuation of the incident and emitted intensity by the absorption in the material.  In general, for a sample that is not infinitely thick or that is inhomogenous along $z$, $\mu(E)$ will depend on $z$.  The intensity is summed over the emission from all elements in the material and all core electrons of those elements, denoted by $X$ ($X$ = Cu $2p$, Cu $3p$, O $1s$, O $2s$, \textellipsis), and $\mu_X(E_i)$ is the contribution to the total linear absorption coefficient from element and core electron $X$. The emission from element and core electron $X$ is proportional to the number of core-holes [i.e., the number of photons absorbed by $X$, $\mu_X(E_i)$] times the fluorescence probability, $\omega_X(E_f,E_i)$ -- the probability that element and core electron $X$ will emit a photon of energy $E_f$ for an incident photon energy $E_i$, as opposed to decaying via Auger or secondary electrons or via emission at a different photon energy.   

For a homogeneous, infinitely thick sample, the integral in Eq.~(\ref{eqn:step1}) simplifies to
\begin{equation}
	 \frac{I(E_i, E_f)}{I_0(E_i)} \!=\! C \sum_{X} \frac{\omega_X(E_i,E_f)\mu_X(E_i)} {\mu(E_i)+\mu(E_f)\frac{\sin \alpha}{\sin \beta}}.
\label{eqn:step3}
\end{equation}
The absorption and subsequent emission can correspond to either exciting a core electron into unoccupied states near threshold (e.g., a Cu $2p$ core electron excited into Cu $3d$ unoccupied states for incident energies at the Cu $L$ absorption edge) or to exciting a core electron well above an absorption edge into the free-electron-like continuum of states (e.g., a O $1s$ electron excited into the continuum for incident energies at the Cu $L$ absorption edge).  The x-ray emission corresponding to these two processes can be characterized as either resonant emission (RXES) or normal emission (NXES).  This distinction is important --  $\omega_X(E_f,E_i)$ and $\mu_X(E_i)$ will be relatively constant functions of $E_i$ for NXES, but will vary strongly for RXES.  

If one scans the incident photon energy through an absorption edge (e.g. Cu $L$) and uses an energy-sensitive detector to measure only the NXES (emission only occurs at discrete emission energies) from element and core electron $Y$ (e.g., O $K$ emission) and not the RXES (e.g., Cu $L$ emission), Eq.~(\ref{eqn:step3}) simplifies to
\begin{equation}
	 \frac{I(E_i, E_f)}{I_0(E_i)} = C \frac{\omega_Y(E_i,E_f)\mu_Y(E_i)} {\mu(E_i)+\mu(E_f)\frac{\sin \alpha}{\sin \beta}}.
\label{eqn:step4}
\end{equation}
At $E_i$, $\omega_Y(E_i,E_f) = \omega_Y(E_f)$ and  $\mu_Y(E_i) = \mu_Y$ are approximately constant, as is $\mu(E_f)$.  A straightforward inversion of Eq.~(\ref{eqn:step4}) defines the IPFY:
\begin{equation}
	\mathrm{IPFY} =  \frac{I_0(E_i)}{I(E_i, E_f)} \!\approx\! A(\mu(E_i)+B) 
\label{eqn:step5}
\end{equation}
where $A = 1/(C \omega_Y(E_f) \mu_Y)$ and $B = \mu(E_f)\frac{\sin \alpha}{\sin \beta} $ are approximately constant.  Equation~(\ref{eqn:step5}) shows that IPFY is linearly proportional to the linear attenuation coefficient, $\mu(E_i)$, without a self-absorption or saturation correction. The offset in the spectra, $AB$, depends on geometry and is minimized for small $\sin\alpha$ (grazing incidence) and/or large $\sin\beta$ (normal detection). This optimal geometry is notably opposite to that of TFY, where self-absorption effects are minimized for large $\sin\alpha/\sin\beta$.\cite{Eisebitt93} The spectra in Fig.~\ref{fig:fig2}(c) show how this offset can manifest itself in IPFY measurements. By subtracting the offset, the spectra measured for different geometries collapse onto a single curve, as shown in Fig.~\ref{fig:fig2}(d). 

\pdfbookmark[1]{Discussion}{Discussion}
Measuring XAS with IPFY should be generally applicable in any instance in which one has a source of normal emission in addition to resonant emission processes.  The source of NXES can be from atoms in the material other than the atom corresponding to the x-ray absorption edge being investigated, as we have shown using the O $K$ emission to provide the Nd $M$ and Cu $L$ edge absorption spectra.  However, one could presumably also use the normal emission from another core electron in the same element as the one investigated at the absorption edge to probe the XAS in a monatomic material, such as using the $N$ emission to probe the $M$ edge absorption in rare-earth materials or the $L$ emission to probe the $K$ edge absorption in the 3$d$ transition metals.  The combinations of absorption edges and emission lines available indicate that this technique should be widely applicable to a range of materials using either hard or soft x-rays.  In particular, IPFY could be used whenever there are issues with surface quality -- where the greater surface sensitivity of TEY may give erroneous results -- and surface treatment is not possible or inconvenient.  In the event of a contaminated surface, the IPFY will be given by Eq.~(\ref{eqn:step5}) provided $\exp(-[\mu_s(E_i)/\sin\alpha + \mu_s(E_f)/\sin\beta]d) \simeq 1$, where $d$ and $\mu_s$ are the thickness and absorption coefficient the contaminated surface, respectively (i.e., for $d$ thin relative to the x-ray penetration depth).  

A limitation of this technique is that it requires a detector with minimal background (dark counts) and sufficient energy resolution to separate the NXES from RXES.  In our test case of the Cu $L$ and Nd $M$ edges of Nd-doped LSCO, both of these criteria are satisfied with a commercially available silicon drift detector which has negligible dark counts and an energy resolution of $\sim$100 eV. Even this crude energy resolution is sufficient to clearly separate the O $K$ NXES from Cu $L$ or Nd $M$ RXES.  However, a similar measurement probing the La $M$ edge (3$d$ to 4$f$) absorption (not shown) in the same material by measuring the O $K$ IPFY provides a distorted spectrum due to finite energy resolution of the detector.  In this case, La 4$s$ to 3$p$ emission ($M_{\zeta}$) at  $\sim$630 eV results in RXES counts at the O $K$ emission energy of $\sim$525 eV due to the finite energy resolution of the detector.  This complication can be circumvented by using detectors with higher energy resolution, such as high-resolution spectrometers used in RXES measurements \cite{KotaniRMP} or, in principle, by using fitting procedures to separate the NXES and RXES.

Our IPFY method differs from past PFY measurements that rely on RXES from the element being investigated by the incident photon energy, either as a means to study dilute concentrations of an element \cite{Jaklevic77} or to provide energy resolution enhanced XAS.\cite{Hamalainen91}  In both instances, as with TFY, self-absorption effects may occur in the spectra since $\mu_X(E_i)$ enters as both the rate at which core holes are created and as a factor in the x-ray penetration depth into the sample [as a contribution to $\mu_{tot}(E_i)$].  If $\mu_X(E_i)$ is a sizable fraction of $\mu_{tot}(E_i)$, the measured spectra will be distorted relative to the object of interest, $\mu(E_i)$.\cite{Eisebitt93, Troger92}  Limited measurements of the NXES PFY have been performed in the past as means to investigate potential multi-atom resonant photoemission effects (MARPEs),\cite{Arenholz00,Moewes00} however, these previous measurements fail to make the connection that the IPFY is proportional to the absorption coefficient.   Furthermore, the agreement between TEY and the IPFY in our samples suggests that the MARPE effect is insignificant, at least at the Cu $L$ and Nd $M$ edges of LNSCO.  

Finally, the energy-resolved fluorescence measurements shown in Fig.~\ref{fig:fig1} highlight an additional difficulty with traditional TFY measurements besides self-absorption effects.  In LNSCO, for example, a sizable contribution to the TFY near the Cu $L$ and Nd $M$ absorption edges is due to normal emission from other elements such as the O $K$ and La $M$ emissions, which dip near the absorption edges.  This can lead to significant distortions of the spectra that may be larger than any self-absorption effects and can complicate the use of self-absorption correction methods for TFY which are nonetheless appropriate for PFY.\cite{Troger92,Eisebitt93}  Furthermore, depending on sample geometry, elastic scattering or specular reflection from the sample can unintentionally contribute to PFY or TFY, but will not contribute to IPFY. 

In conclusion, we have demonstrated IPFY as an alternative measure of XAS that can be both bulk sensitive and free of self-absorption effects. IPFY uses normal x-ray emission from an element or core-hole excitation that is different than the x-ray absorption edge being probed.  The inverse of the emission spectra is found to be proportional to the x-ray absorption coefficient, $\mu(E_i)$.  This technique should be generally applicable to a wide range of x-ray absorption measurements using either hard or soft x-rays.  

\begin{acknowledgments}
We gratefully thank G. A. Sawatzky for valuable discussions. This work was supported by NSERC. The research described in this paper was performed at the Canadian Light Source, which is supported by NSERC, NRC, CIHR, and the University of Saskatchewan.  
\end{acknowledgments}

\bibliography{IPFYtransfer}

\end{document}